\documentclass[aps,showpacs,preprintnumbers,amsmath,amssymb]{revtex4}
\usepackage{dcolumn}
\usepackage[dvips]{epsfig}

\begin{document}
\baselineskip=0.8 cm
\title{\bf Dynamical evolution of the electromagnetic perturbation with Weyl corrections}

\author{Songbai Chen\footnote{csb3752@163.com}, Jiliang Jing
\footnote{jljing@hunnu.edu.cn}}
%\email{csb3752@163.com}

\affiliation{Institute of Physics and Department of Physics, Hunan
Normal University,  Changsha, Hunan 410081, People's Republic of
China \\ Key Laboratory of Low Dimensional Quantum Structures \\
and Quantum Control of Ministry of Education, Hunan Normal
University, Changsha, Hunan 410081, People's Republic of China\\
Kavli Institute for Theoretical Physics China, CAS, Beijing 100190,China}

\begin{abstract}
\baselineskip=0.6 cm
\begin{center}
{\bf Abstract}
\end{center}

We present firstly the master equation of an electromagnetic perturbation with Weyl correction in the four-dimensional black hole spacetime, which depends not only on the Weyl correction parameter, but also on the parity of the electromagnetic field. It is quite different from that of the usual electromagnetic perturbation without Weyl correction in the four-dimensional spacetime. And then we have investigated numerically the dynamical evolution of the electromagnetic perturbation with Weyl correction in the background of
a four-dimensional Schwarzschild black hole spacetime. Our results show that the Weyl correction parameter $\alpha$ and the parities imprint in the wave dynamics of the electromagnetic perturbation.
For the odd parity electromagnetic perturbation, we find it grows with exponential rate if the value of $\alpha$ is below the negative critical value $\alpha_c$. However, for the electromagnetic perturbation with even parity, we find that there does not exist such a critical threshold value and the electromagnetic field always decays in the allowed range of $\alpha$.

\end{abstract}

 \maketitle
\newpage
\section{Introduction}

For the last few decades, one of most interesting topics is to
study the dynamical evolution of an external perturbation around a black hole.
It is widely believed that the quasinormal modes in the
dynamical evolution carry the characteristic information about the
black hole and  could help us to
identify whether there exists black hole in our Universe or not
\cite{Chandrasekhar:1975,Nollert:1999,Kokkotas:1999}. The further investigations
imply that the quasinormal spectrum of black holes could open a window for us to
understand more deeply about the quantum gravity
\cite{Hod:1998,Dreyer:2003,Corichi:2003} and the AdS/CFT
correspondence \cite{Maldacena:1998,Witten:1998,Horowitz:2000}.
Moreover, the dynamical behaviors of the perturbations could be used to test the
stability of a black hole in various theories of gravity
\cite{Gregory:1993,Harmark:2007,Konoplya:2008,Chen:2009}.
The dynamical evolution of various perturbations have been studied extensively in
the various black holes spacetime
\cite{Price:1972,Hod:1998ja,Burko2003,Koyama:2001,Hod:1998ma,Cardoso2001,Barack:2000}.

All of the above investigations mentioned for the electromagnetic
perturbation are in the frame of Einstein-Maxwell electromagnetic theory in which the Maxwell Lagrangian is only quadratic
in the Maxwell tensor and does not contain any coupling between the Maxwell part and the curvature part. Recently, a lot of attention have been focused on studying the generalized Einstein-Maxwell theory. The main motivation is
that the generalized Einstein-Maxwell theory contains higher
derivative interactions and carries more information about the electromagnetic field. The study of the generalized Einstein-Maxwell theory could help us to
explore the full properties and effects of the electromagnetic fields. One of interesting generalized Einstein-Maxwell theory is Born-Infeld theory which is introduced in the thirties \cite{Born} in order to remove the divergence of the electron's self-energy in the classical electrodynamics. Moreover,
Born-Infeld theory displays good physical properties concerning wave propagation such as the absence of shock waves and birefringence phenomena
\cite{Boillat}. Born-Infeld theory has also received special
attention because it could arise in the low-energy regime of string
and D-Brane physics \cite{Fradkin}.  Another generalization of the Einstein-Maxwell theory with three parameters has been considered in which there are the non-minimal
couplings between the gravitational and electromagnetic fields in the Lagrangian \cite{Balakin,Faraoni,Hehl,Balakin1}. The presence of such non-minimal
couplings in the Lagrangian modifies of the coefficients
involving the second-order derivatives both in the Maxwell and Einstein equations, which could affect the propagation of gravitational and electromagnetic waves in the spacetime and may yield time delays in the arrival of those waves \cite{Balakin}. Moreover, these couplings could modify the electromagnetic and gravitational structure of a charged black hole \cite{Balakin1}. In the evolution of the early Universe, these coupled terms may yield electromagnetic quantum fluctuations and lead to the inflation  \cite{Turner,Mazzitelli,Lambiase,Raya,Campanelli}. Due to the inflation at that time, the scale of the fluctuations can be stretched towards outside the Hubble horizon and then they result in classical fluctuations, which means that the non-minimal couplings could be used to explain
the large scale magnetic fields observed in clusters of galaxies \cite{Bamba,Kim,Clarke}.

In this paper, we consider a simple
generalized electromagnetic theory which involves a coupling between
the Maxwell field and the Weyl tensor \cite{Weyl1,Drummond}.
In this theory, the Lagrangian density of the electromagnetic
field is modified as
\begin{eqnarray}\label{LEM}
L_{EM}=-\frac{1}{4}\bigg(F_{\mu\nu}F^{\mu\nu}-4\alpha
C^{\mu\nu\rho\sigma}F_{\mu\nu}F_{\rho\sigma}\bigg),
\end{eqnarray}
where $C_{\mu\nu\rho\sigma}$ is the Weyl tensor and $\alpha$ is a
coupling constant with dimensions of length-squared. $F_{\mu\nu}$ is
the electromagnetic tensor, which is related to the electromagnetic
vector potential $A_{\mu}$ by $F_{\mu\nu}=A_{\nu;\mu}-A_{\mu;\nu}$.
Actually, the coupling in the Lagrangian density (\ref{LEM}) is a special of coupling between the gravitational and electromagnetic fields
since Weyl tensor $C_{\mu\nu\rho\sigma}$ is related to the Riemann tensor $R_{\mu\nu\rho\sigma}$, the Ricci tensor $R_{\mu\nu}$ and the Ricci scalar $R$ by
\begin{eqnarray}
C_{\mu\nu\rho\sigma}=R_{\mu\nu\rho\sigma}-\frac{2}{n-2}(
g_{\mu[\rho}R_{\sigma]\nu}-g_{\nu[\rho}R_{\sigma]\mu})+\frac{2}{(n-1)(n-2)}R
g_{\mu[\rho}g_{\sigma]\nu},
\end{eqnarray}
where $n$ and $g_{\mu\nu}$ are the dimension and metric of the spacetime, and brackets around indices refers to the antisymmetric part.
It was found that the similar couplings between curvature tensor and Maxwell
tensor could be obtained from a calculation in QED of the photon
effective action from one-loop vacuum polarization on a curved
background \cite{Drummond}. Although Weyl correction can  be looked as an effective description of quantum effects, such kind of couplings may also occur near classical compact astrophysical objects with high mass density and strong gravitational field such as the supermassive black holes at the center of galaxies \cite{Dereli1}. The effects originating from such kind of couplings could also be used to distinguish between general relativity and other theories of gravity in the future astrophysical observations \cite{Solanki}.
Therefore, we can treat the Weyl correction to electromagnetic field as a kind of general classical couplings between the gravitational and electromagnetic fields and study the effects of such correction on the dynamical evolution of electromagnetic field in the general background spacetime.
The
coupling term with Weyl tensor is a tensorial structure correcting
the Maxwell term at leading order in derivatives, which modifies the
Einstein-Maxwell equation and affects the dynamical evolution of
electromagnetic field in the background spacetime. An advantage of
this generalized electromagnetic theory with Weyl correction is that
the modified Einstein-Maxwell equation is not complicated and the
equations of motion for electromagnetic perturbation can be
decoupled to a second order differential equation, which is very
important for us to investigate further the dynamical properties of
electromagnetic perturbation in the black hole spacetime. In
Ref.\cite{Weyl1}, the authors studied the holographic conductivity
and charge diffusion with Weyl correction in the anti-de Sitter
(AdS) spacetime and found that the correction breaks the universal
relation with the $U(1)$ central charge observed at leading order.
Recently, the holographic superconductors with Weyl corrections are
also explored in \cite{Wu2011,Ma2011,Momeni,Roychowdhury}. Wu
\textit{et al} \cite{Wu2011} studied the effects of Weyl corrections
$s$-wave holographic superconductor and found that with Weyl
corrections the critical temperature becomes smaller and the scalar
hair is formed harder when the coupling constant $\alpha$ is
negative, but the result is just opposite when the constant $\alpha$
is positive. In the St\"{u}ckelberg mechanism, it is found that Weyl
coupling parameter $\alpha$ also changes the order of the phase
transition of the holographic superconductor \cite{Ma2011}. The
$p$-wave holographic superconductor model with Weyl corrections has
been studied and it is shown that the effect of Weyl corrections on
the condensation is similar to that of the $s$-wave model
\cite{Momeni}. Moreover, the effects of Weyl corrections on the
phase transition between the holographic insulator and
superconductor has been investigated in \cite{zhao2013} and it is
found that in this case the effects of Weyl corrections depend on
the model of holographic dual. For the $p$-wave model, the higher
Weyl corrections will make it harder for the holographic
insulator/superconductor phase transition to be triggered. However,
for the $s$-wave model, the Weyl couplings do not affect the
properties of the holographic insulator/superconductor phase
transition since the critical chemical potentials are independent of
the Weyl correction terms in this case. These results may excite
more efforts to be focused on the study of the electrodynamics with
Weyl corrections in the more general cases. The main purpose of this
paper is to investigate the dynamical evolution of the
electromagnetic perturbation coupling to the Weyl tensor in the
Schwarzschild black hole spacetime and see the effect of the Weyl
corrections on the stability of the black hole.

The plan of our paper is organized as follows: in the following
section we will derive the master equation of electromagnetic perturbation
with Weyl correction in the four-dimensional static and spherical
symmetric spacetime. In Sec.III, we will
study numerically the effects of the Weyl corrections on the quasinormal modes of the
electromagnetic perturbation in the Schwarzschild black hole and then examine
the stability of the black hole. Finally, in the last section we will include
our conclusions.

\section{The wave equation for the electromagnetic perturbations with Weyl corrections}

In order to study the effects of Weyl corrections on the dynamical
evolution of the electromagnetic perturbations in a black hole
spacetime, we must first obtain its wave equation in the background.
The action of Maxwell field with Weyl corrections in the curved
spacetime has a form \cite{Weyl1}
\begin{eqnarray}
S=\int d^4x \sqrt{-g}\bigg[\frac{R}{16\pi
G}-\frac{1}{4}\bigg(F_{\mu\nu}F^{\mu\nu}-4\alpha
C^{\mu\nu\rho\sigma}F_{\mu\nu}F_{\rho\sigma}\bigg)\bigg].\label{acts}
\end{eqnarray}
Varying the action (\ref{acts}) with respect to $A_{\mu}$, one can
obtain the generalized Maxwell equation
\begin{eqnarray}
\nabla_{\mu}\bigg(F^{\mu\nu}-4\alpha
C^{\mu\nu\rho\sigma}F_{\rho\sigma}\bigg)=0.\label{WE}
\end{eqnarray}
Obviously, the Weyl corrections affect the dynamical evolution of
the electromagnetic perturbation.

For a four-dimensional static and spherical
symmetric black hole spacetime, the metric has a form
\begin{eqnarray}
ds^2&=&fdt^2-\frac{1}{f}dr^2-r^2
d\theta^2-r^2\sin^2{\theta}d\phi^2,\label{m1}
\end{eqnarray}
where the metric coefficient $f$ is a function of polar coordinate $r$. In this background,
one can expand $A_{\mu}$ in vector spherical harmonics \cite{Ruffini}
\begin{eqnarray}
A_{\mu}= \sum_{l,m}\left(\left[\begin{array}{ccc}
 &0&\\
 &0&\\
 &\frac{a^{lm}(t,r)}{\sin\theta}\partial_{\phi}Y_{lm}&\\
 &-a^{lm}(t,r)\sin\theta\partial_{\theta}Y_{lm}&
\end{array}\right]+\left[\begin{array}{cccc}
 &j^{lm}(t,r)Y_{lm}&\\
 &h^{lm}(t,r)Y_{lm}&\\
 &k^{lm}(t,r)\partial_{\theta}Y_{lm}&\\
 &k^{lm}(t,r)\partial_{\phi}Y_{lm}&
\end{array}\right]\right),\label{Au}
\end{eqnarray}
where the first term in the right side has parity $(-1)^{l+1}$ and
the second term has parity $(-1)^{l}$, $l$ is the angular quantum
number and $m$ is the azimuthal number.

Adopting the following form
\begin{eqnarray}
a^{lm}(t,r)&=&a^{lm}(r)e^{-i\omega t},~~~~~h^{lm}(t,r)=h^{lm}(r)e^{-i\omega t},\nonumber\\
j^{lm}(t,r)&=&j^{lm}(r)e^{-i\omega t},~~~~~k^{lm}(t,r)=k^{lm}(r)e^{-i\omega t},
\end{eqnarray}
and then inserting the above expansion (\ref{Au})
into the generalized Maxwell equation (\ref{WE}), we can obtain three independent coupled differential equations. Eliminating $k^{lm}(r)$, we can
get a
second order differential equation for the perturbation
\begin{eqnarray}
\frac{d^2\Psi(r)}{dr^2_*}+[\omega^2-V(r)]\Psi(r)=0,\label{radial}
\end{eqnarray}
where the tortoise coordinate $r_{*}$ is defined as $dr_*=\frac{
dr}{f}$. The wavefunction $\Psi(r)$ is a linear combination of the
functions $j^{lm}$, $h^{lm}$, and $a^{lm}$, which appeared
in the expansion (\ref{Au}). The form of $\Psi(r)$ depends on the
parity of the perturbation, which can be expressed as
\begin{eqnarray}
\Psi(r)=a^{lm}\sqrt{1+\frac{2\alpha}{3r^2}(r^2f''-2rf'+2f-2)},
\end{eqnarray}
for the odd parity $(-1)^{l+1}$, and
\begin{eqnarray}
\Psi(r)=\frac{r^2}{l(l+1)}\bigg(-i\omega
h^{lm}-\frac{dj^{lm}}{dr}\bigg)
\frac{\sqrt{1-\frac{2\alpha}{3r^2}(r^2f''-2rf'+2f-2)}}{
1-\frac{4\alpha}{3r^2}(r^2f''-2rf'+2f-2)},
\end{eqnarray}
for the even parity $(-1)^{l}$, respectively. The effective
potential $V(r)$ in Eq. (\ref{radial}) depends also on the parity of
the perturbation. For the odd parity, the potential $V(r)$ is given
by
\begin{eqnarray}
V(r)=f\bigg\{\frac{l(l+1)}{r^2}\frac{1-\frac{4\alpha}{3r^2}(r^2f''-2rf'+2f-2)}{
1+\frac{2\alpha}{3r^2}(r^2f''-2rf'+2f-2)}+\frac{\alpha(h_0+h_1\alpha)}{r^2[
3r^2+2\alpha(r^2f''-2rf'+2f-2)]^2}\bigg\},\label{evod}
\end{eqnarray}
whereas for the even parity it is given by
\begin{eqnarray}
V(r)=f\bigg\{\frac{l(l+1)}{r^2}\frac{1+\frac{2\alpha}{3r^2}(r^2f''-2rf'+2f-2)}{
1-\frac{4\alpha}{3r^2}(r^2f''-2rf'+2f-2)}-\frac{\alpha(h_0+h_2\alpha)}{r^2[
3r^2+2\alpha(r^2f''-2rf'+2f-2)]^2}\bigg\},\label{even}
\end{eqnarray}
where
\begin{eqnarray}
h_0&=&3r^2[12f^2+rf'(r^3f^{(3)}-2r^2f''+4rf'+4)+f(r^4f^{(4)}-2r^3f^{(3)}+6r^2f''-16rf'-12)],\\
h_1&=&32f^3+2rf'(r^2f''-2rf'-2)(r^3f^{(3)}-2r^2f''+4rf'+4)+4f^2(r^4f^{(4)}+8r^2f''-20rf'-16)\nonumber\\
&-&f[4r^4f^{(4)}+(r^3f^{(3)})^2-8r^4f''^{2}-64r^2f'^2-32+4rf'(r^4f^{(4)}-r^3f^{(3)}+12r^4f''-24)\nonumber\\
&+&2r^2f''(16-r^4f^{(4)})],
\\
h_2&=&-16r^3f'^3-4r^2f'^2(r^3f^{(3)}-4r^2f''-8rf'+8)+r^2f[2(4f^{(3)}+rf^{(4)})(r^2f''+2f-2)-3r^3(f^{(3)})^2]\nonumber\\
&-&2f'[8f^2-(r^2f''-2)(r^3f^{(3)}-2r^2f''+4)+2f(r^4f^{(4)}+3r^3f^{(3)}+4r^2f''-8)].
\end{eqnarray}
It is clear that the coupling constant $\alpha$ emerges in the
effective potential, which means that Wely corrections will change
the dynamical evolution of the electromagnetic perturbation in the
background spacetime. From Eqs. (\ref{evod}) and (\ref{even}), we
also find that the modification of the effective potential by Weyl
corrections is different for the electromagnetic perturbations with
different parities, which implies that the effects of Weyl
corrections on the wave dynamics for the electromagnetic
perturbation with the odd parity are different from those of the
perturbation with the even parity. This is quite different from that
in the case of the usual electromagnetic perturbation without Weyl
corrections in the four-dimensional spacetime, in which the electromagnetic perturbation with the odd
parity has the same effective potential as for the perturbation with
the even parity. When the coupling
constant $\alpha=0$ the effective potentials (\ref{evod}) and
(\ref{even}) recover to the usual form without Weyl corrections.

\section{Effects of Weyl corrections on the wave dynamics of the electromagnetic perturbation in the Schwarzschild black hole spacetime}

In this section, we will study numerically the dynamical evolution of the
electromagnetic perturbation with Weyl corrections in the
background of a Schwarzschild black hole spacetime and probe the
effects of Weyl corrections on the wave dynamics of the
electromagnetic perturbation.

For the Schwarzschild black hole spacetime, the metric function is
$f=1-\frac{2M}{r}$ and then the effective potentials (\ref{evod}) and
(\ref{even}) can be expressed as
\begin{eqnarray}
V(r)_{odd}=(1-\frac{2M}{r})\bigg[\frac{l(l+1)}{r^2}(\frac{r^3+16\alpha
M}{r^3-8\alpha M})-\frac{24\alpha M(2r^4-5Mr^3-10\alpha Mr+28\alpha
M^2)}{r^3 (r^3-8\alpha M)^2}\bigg],\label{evodsch}
\end{eqnarray}
for the odd parity and
\begin{eqnarray}
V(r)_{even}=(1-\frac{2M}{r})\bigg[\frac{l(l+1)}{r^2}(\frac{r^3-8\alpha
M}{r^3+16\alpha M})+\frac{24\alpha M(2r^4-5Mr^3+2\alpha Mr+4\alpha
M^2)}{r^3 (r^3-8\alpha M)^2}\bigg],\label{evensch}
\end{eqnarray}
for the even parity, respectively. Defining the quantity
\begin{eqnarray}
W&=&\frac{12\alpha M(r-2M)}{r^2(r^3-8\alpha
M)}+\frac{3l(l+1)}{4r}+\frac{\sqrt{3}l(l+1)}{16\alpha^3M^3}\bigg[4\arctan\bigg(\frac{r+(\alpha
M)^{\frac{1}{3}}}{\sqrt{3}(\alpha
M)^{\frac{1}{3}}}\bigg)-2^{\frac{2}{3}}\arctan\bigg(\frac{2^{\frac{1}{3}}r-2(\alpha
M)^{\frac{1}{3}}}{2\sqrt{3}(\alpha
M)^{\frac{1}{3}}}\bigg)\bigg]\nonumber\\&+&\frac{l(l+1)}{32\alpha^3M^3}\bigg\{4\log\bigg[\frac{(r-2(\alpha
M)^{\frac{1}{3}})^2}{r^2+2(\alpha M)^{\frac{1}{3}}r+4(\alpha
M)^{\frac{2}{3}}}\bigg]+2^{\frac{2}{3}}\log\bigg[\frac{(2^{\frac{2}{3}}r+4(\alpha
M)^{\frac{1}{3}})^2}{2^{\frac{1}{3}}r^2-2^{\frac{5}{3}}(\alpha
M)^{\frac{1}{3}}r+8(\alpha M)^{\frac{2}{3}}}\bigg]\bigg\},
\end{eqnarray}
we can obtain
\begin{eqnarray}
V(r)_{odd}=W^2+\frac{dW}{dr_*}+\beta,~~~~~~~~~~V(r)_{even}=W^2-\frac{dW}{dr_*}+\beta,
\end{eqnarray}
with
\begin{eqnarray}
\beta=\frac{(r-2M)}{r^3(r^3-8\alpha M)}\bigg[\frac{r^6+8\alpha
M+160\alpha^2 M^2}{(r^3+16\alpha M)}+\frac{144\alpha^2
M^2(r-2M)}{r(r^3-8\alpha M)}\bigg]-W^2.
\end{eqnarray}
This means that these two effective potentials for odd and even
parities can be written in the form of super-partner potentials. The
similar behavior of effective potentials for gravitational
perturbations have been discovered in
\cite{Chandrasekhar1983,Cooper1995}.
\begin{figure}[ht]
\begin{center}
\includegraphics[width=5.5cm]{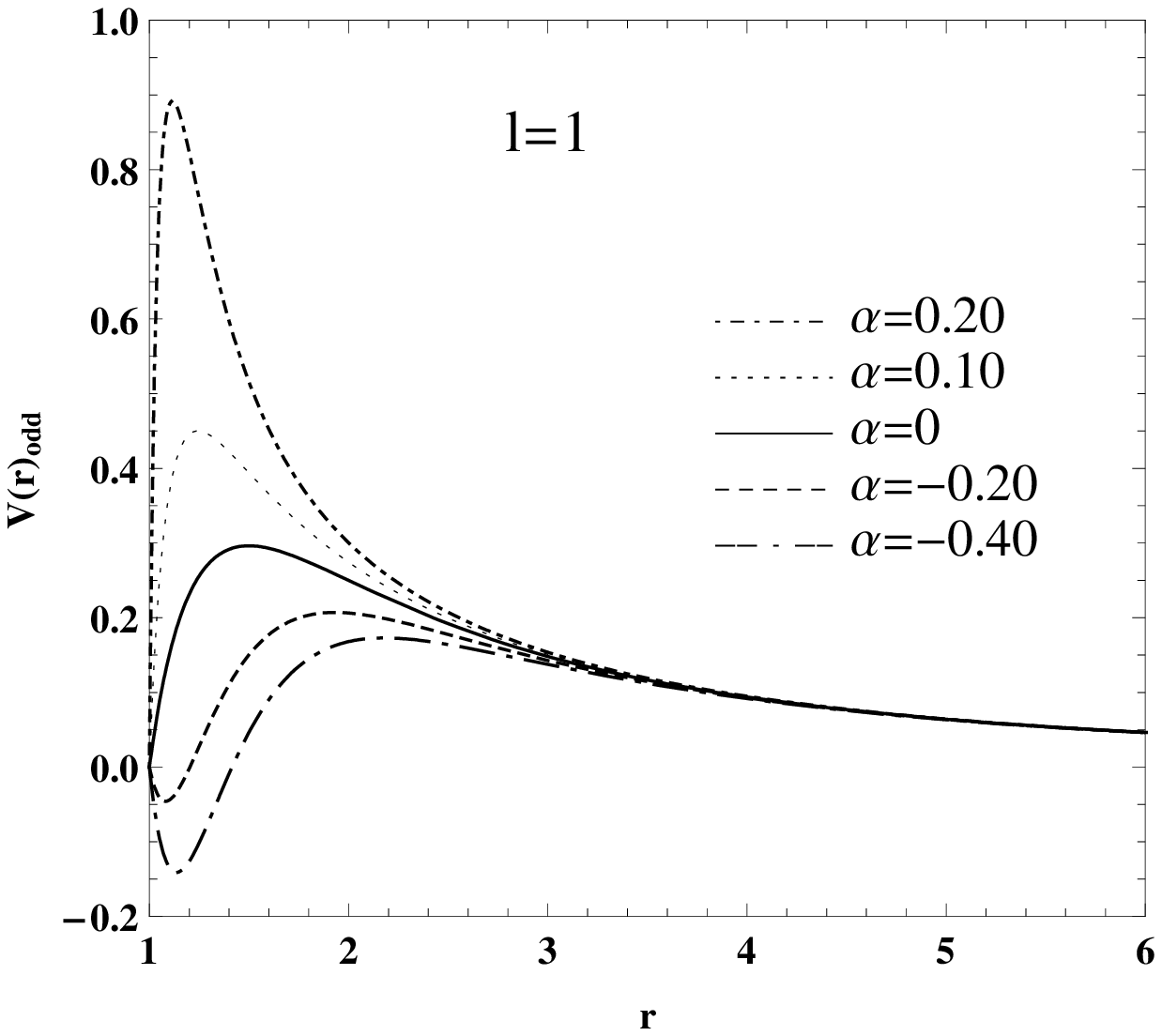}
\includegraphics[width=5.5cm]{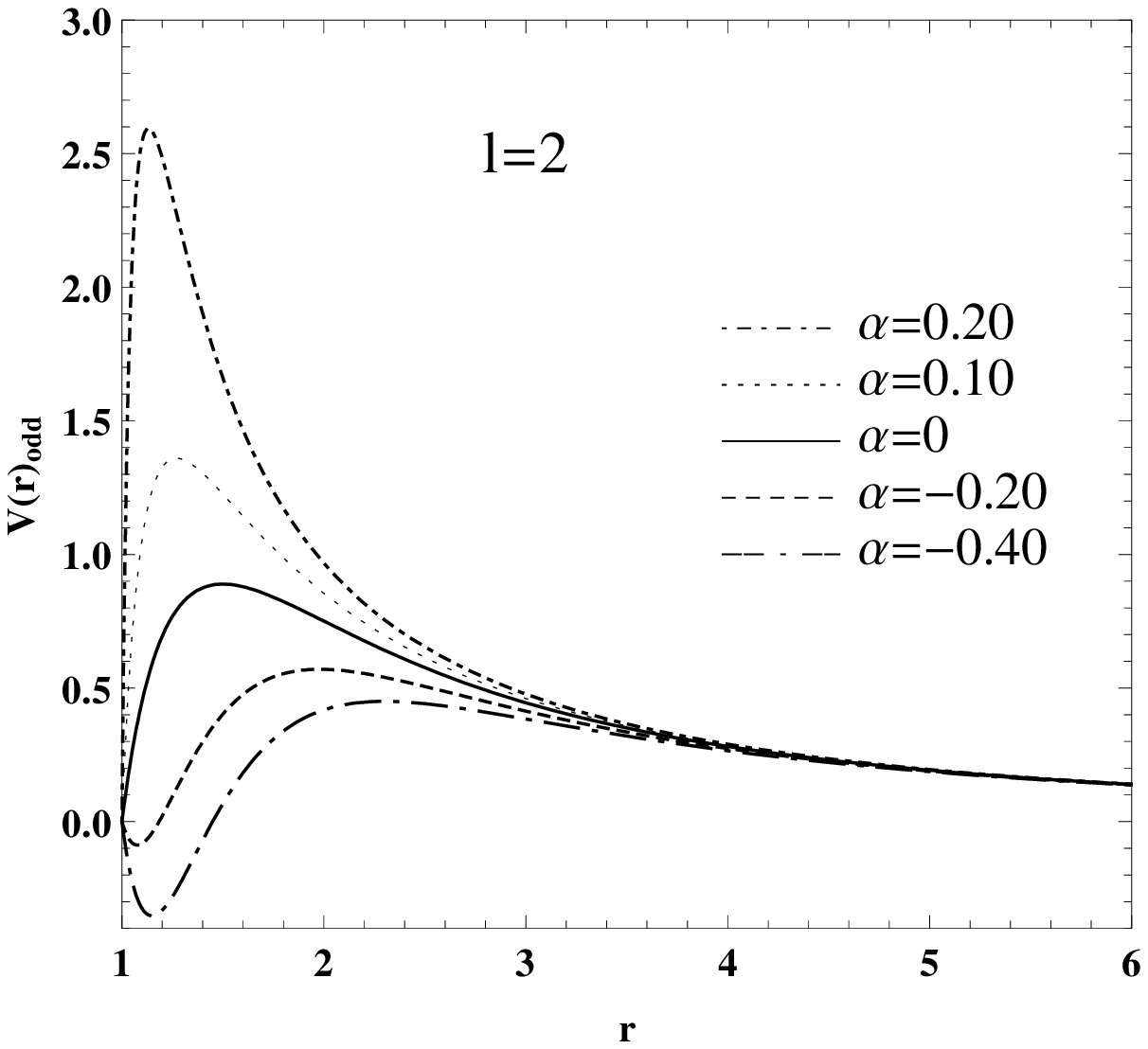}\includegraphics[width=5.6cm]{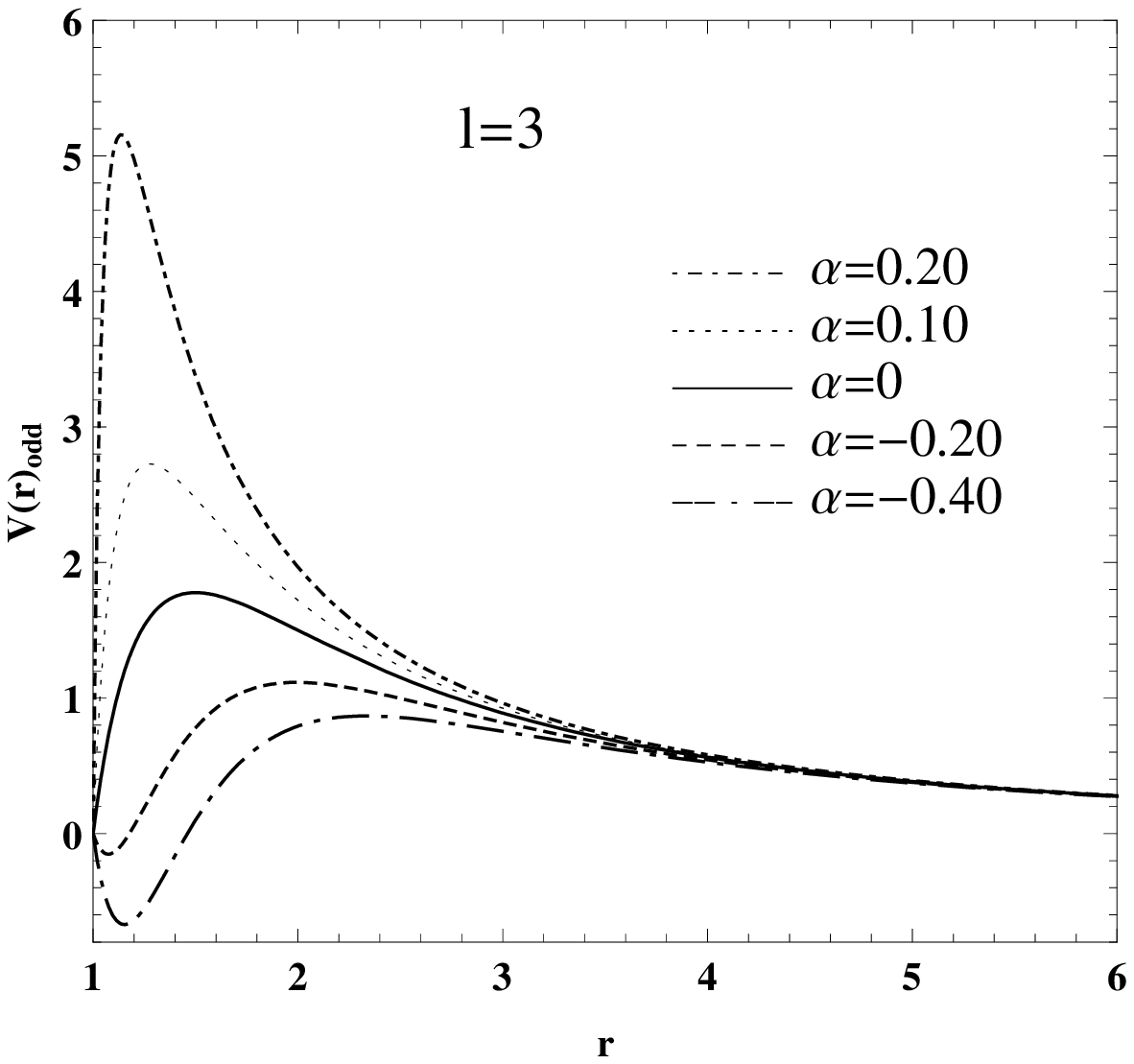}
\caption{Variation of the effective potential $V(r)_{odd}$ with the
polar coordinate $r$ for fixed $l=1$ (left), $l=2$ (middle) and
$l=3$ (right). The long-dash-dotted, dashed, solid, dotted  and
short-dash-dotted lines are corresponding to the cases with
$\alpha=-0.4,\;-0.2,\;0,\;0.1\;0.2$, respectively. We set $2M=1$.}
\end{center}
\end{figure}
\begin{figure}[ht]
\begin{center}
\includegraphics[width=5.5cm]{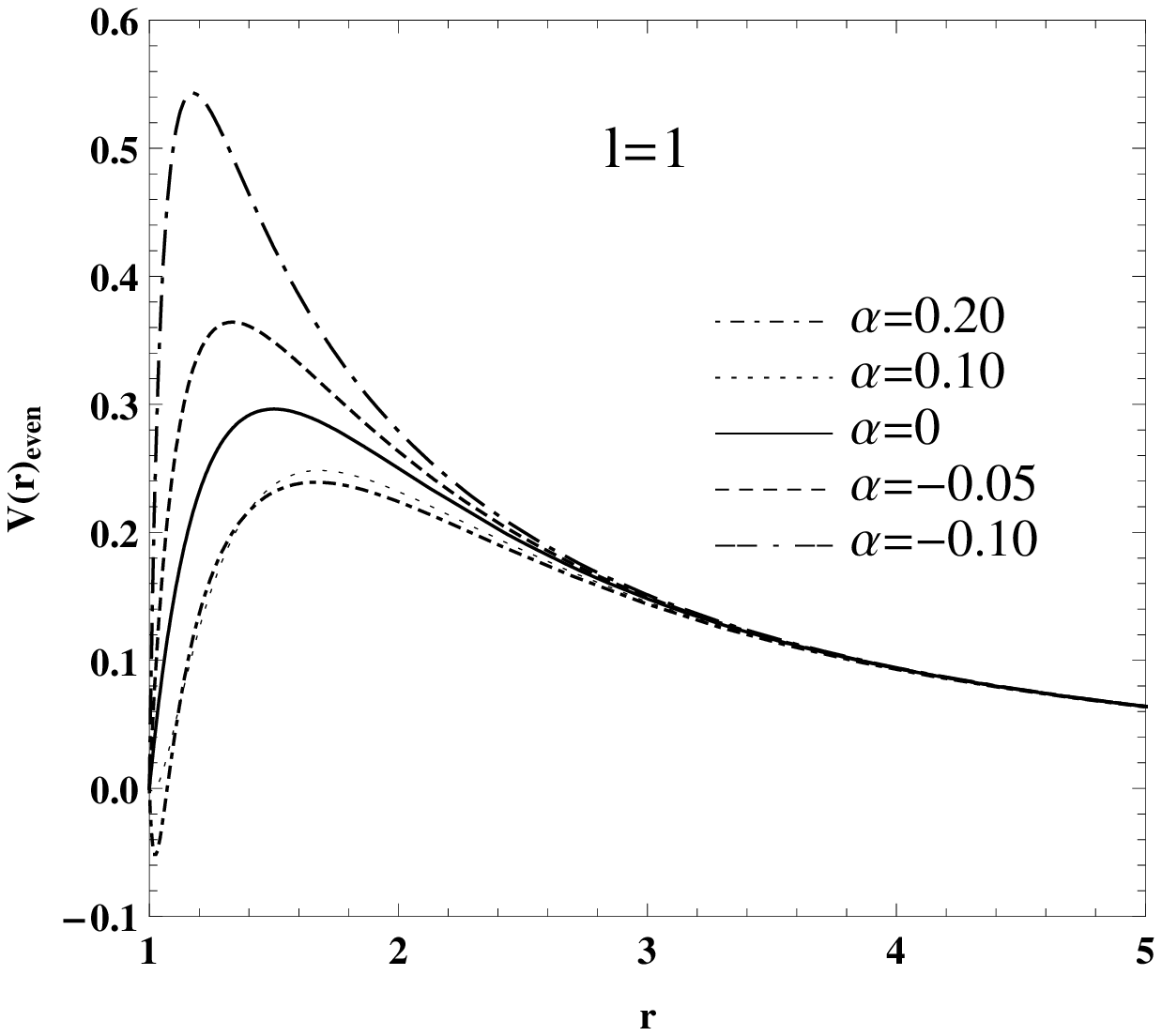}
\includegraphics[width=5.5cm]{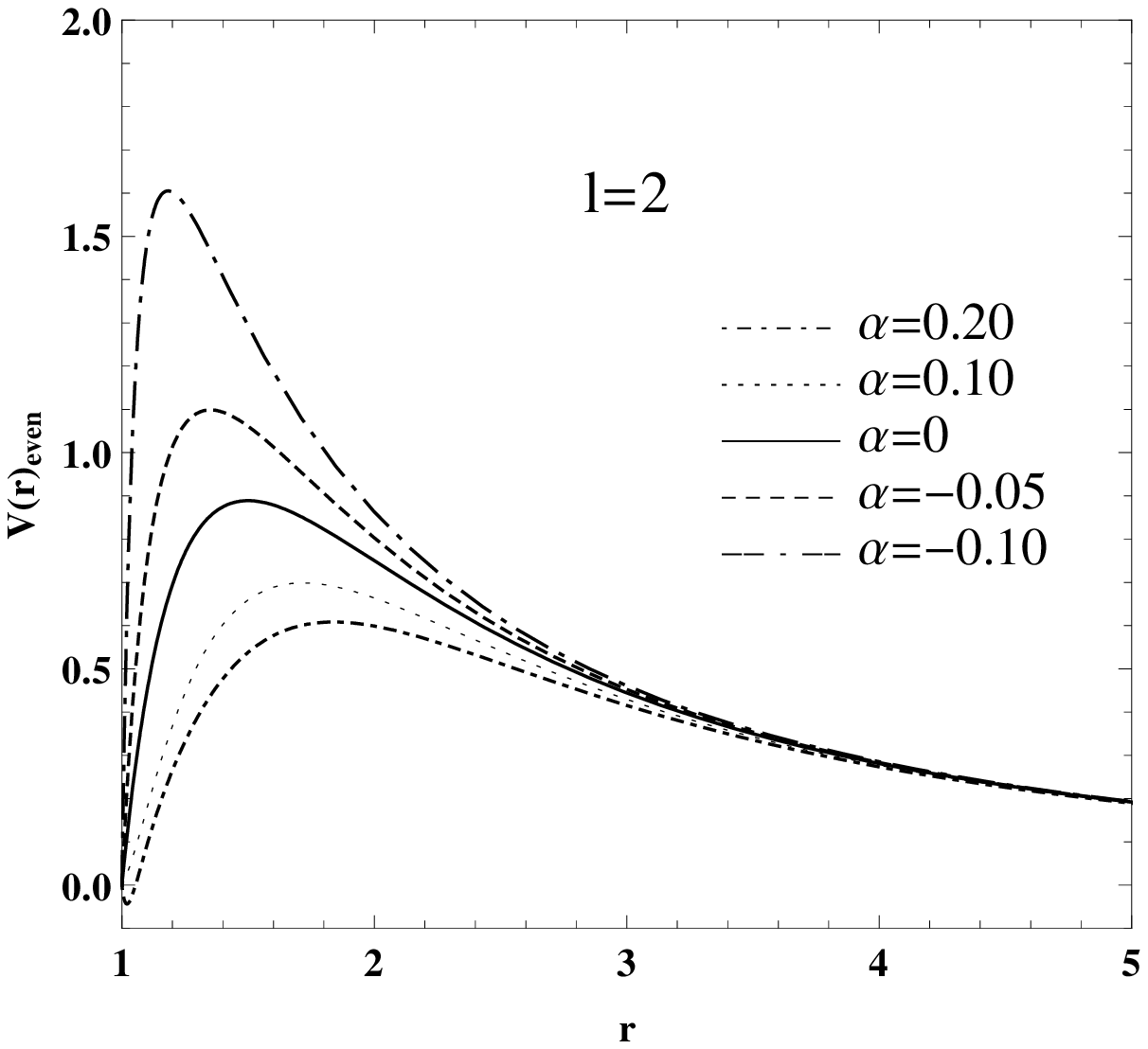}\includegraphics[width=5.6cm]{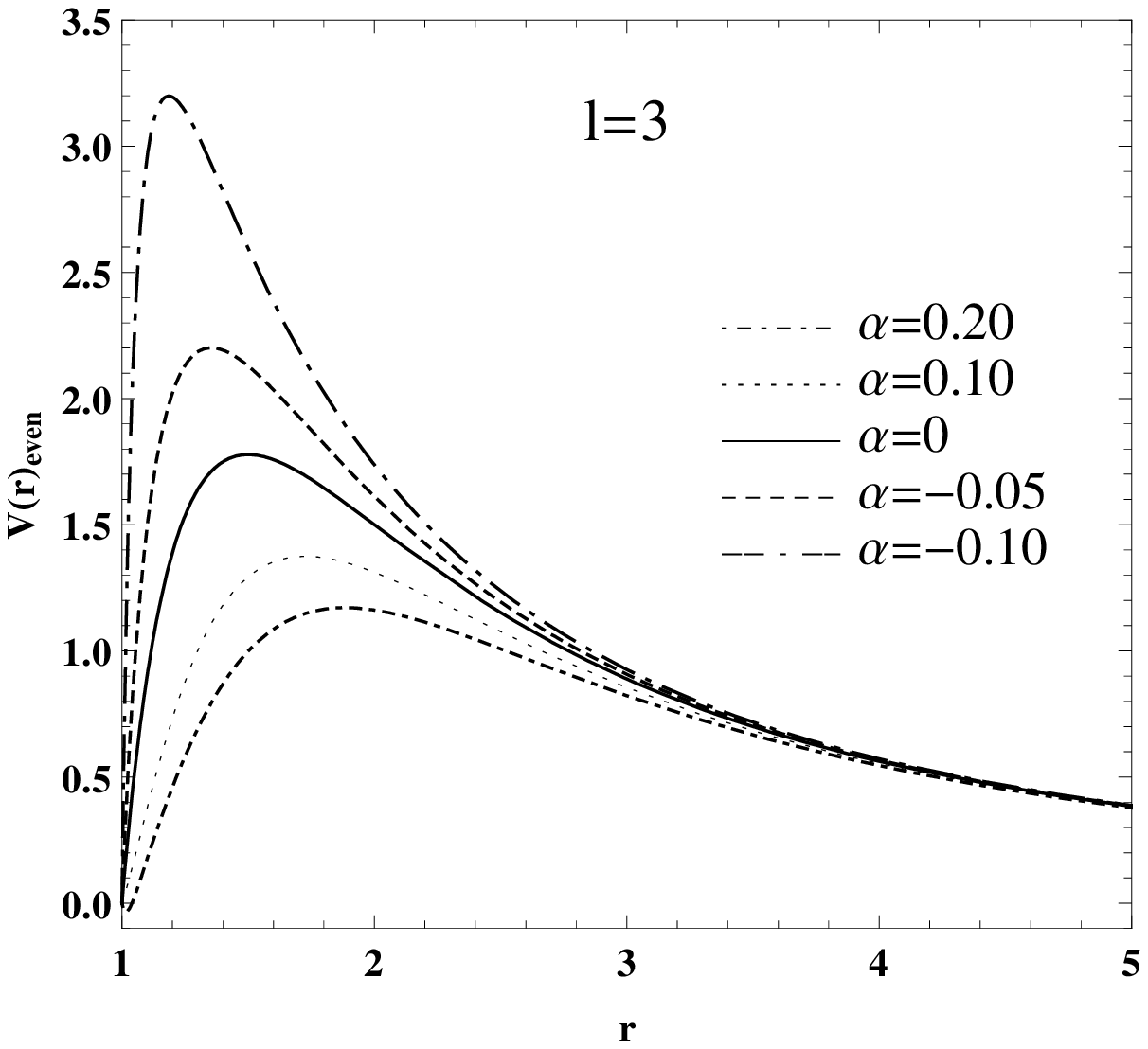}
\caption{Variation of the effective potential $V(r)_{even}$ with the
polar coordinate $r$ for fixed $l=1$ (left), $l=2$ (middle) and
$l=3$ (right). The long-dash-dotted, dashed, solid, dotted  and
short-dash-dotted lines are corresponding to the cases with
$\alpha=-0.1,\;-0.05,\;0,\;0.1\;0.2$, respectively. We set $2M=1$.}
\end{center}
\end{figure}
Considering that the effective potential should be continuous in the
region outside the event horizon of black hole, the coupling
constant $\alpha$ must satisfy $r^3_H-8\alpha M>0$ (i.e.,
$\alpha<M^2$) for the electromagnetic perturbation with the odd
parity, and satisfy $r^3_H-8\alpha M>0$  and $r^3_H+16\alpha M>0$
(i.e., $-\frac{M^2}{2}<\alpha<M^2$) for the perturbation with the
even parity. In Figs.(1) and (2), we show the changes of the
effective potentials $V(r)_{odd}$ and $V(r)_{even}$ with the
coupling constant $\alpha$ for fixed $l$, respectively.  For fixed
$l$, the peak height of the potential barrier increases with the
coupling constant $\alpha$ for $V(r)_{odd}$ and decreases for
$V(r)_{even}$. Moreover, we also find that there exits the negative
gap in the effective potential $V(r)_{odd}$ only for the certain
negative value of $\alpha$ and in the potential $V(r)_{even}$ only
for the certain positive value of $\alpha$. This means that the
properties of the wave dynamics of the electromagnetic perturbation
with the odd parity could be different from those of that with the even
parity. In the following section, we will check those values of
$\alpha$ for which the negative gap is present and study the
stability of the black hole under the electromagnetic perturbation
with Weyl corrections.

Let us now to study the effects of the Weyl corrections on the
quasinormal modes of electromagnetic perturbation in the
Schwarzschild black hole spacetime. In Fig. (3) and (4), we present
the fundamental quasinormal modes ($n=0$) evaluated by the
third-order WKB approximation method \cite{Schutz:1985,Iyer:1987}.
It is shown that with the increase of the coupling parameter $\alpha$
the real parts of the quasinormal frequencies increase for the
electromagnetic perturbation with the odd parity, but decrease for
that with the even parity. The changes of the imaginary parts with
$\alpha$ are more complicated. For the odd parity electromagnetic perturbation, we find that as $\alpha<0$ the imaginary part of  increases with $\alpha$. When $\alpha>0$, the imaginary part for $l=1$
first decreases to the minimum and then rises to the maximum and subsequently decreases to another minimum, and finally it increases
again with $\alpha$, while for other $l$ it first decreases and then increases. For the even parity electromagnetic perturbation,  we find that with the increase of $\alpha$ the imaginary part of first decreases and then increases as $\alpha>0$. When $\alpha<0$, with the increase of $\alpha$ the imaginary part for $l=1$
first decreases to the minimum and then rises to the maximum and subsequently decreases to another minimum, and finally it increases
again to another maximum, while for other $l$ it first decreases and then increases. These results imply that Weyl corrections
modify the standard results of the quasinormal modes for the
electromagnetic perturbations in the background of a Schwarzschild
black hole.

\begin{figure}[ht]
\begin{center}
\includegraphics[width=5.6cm]{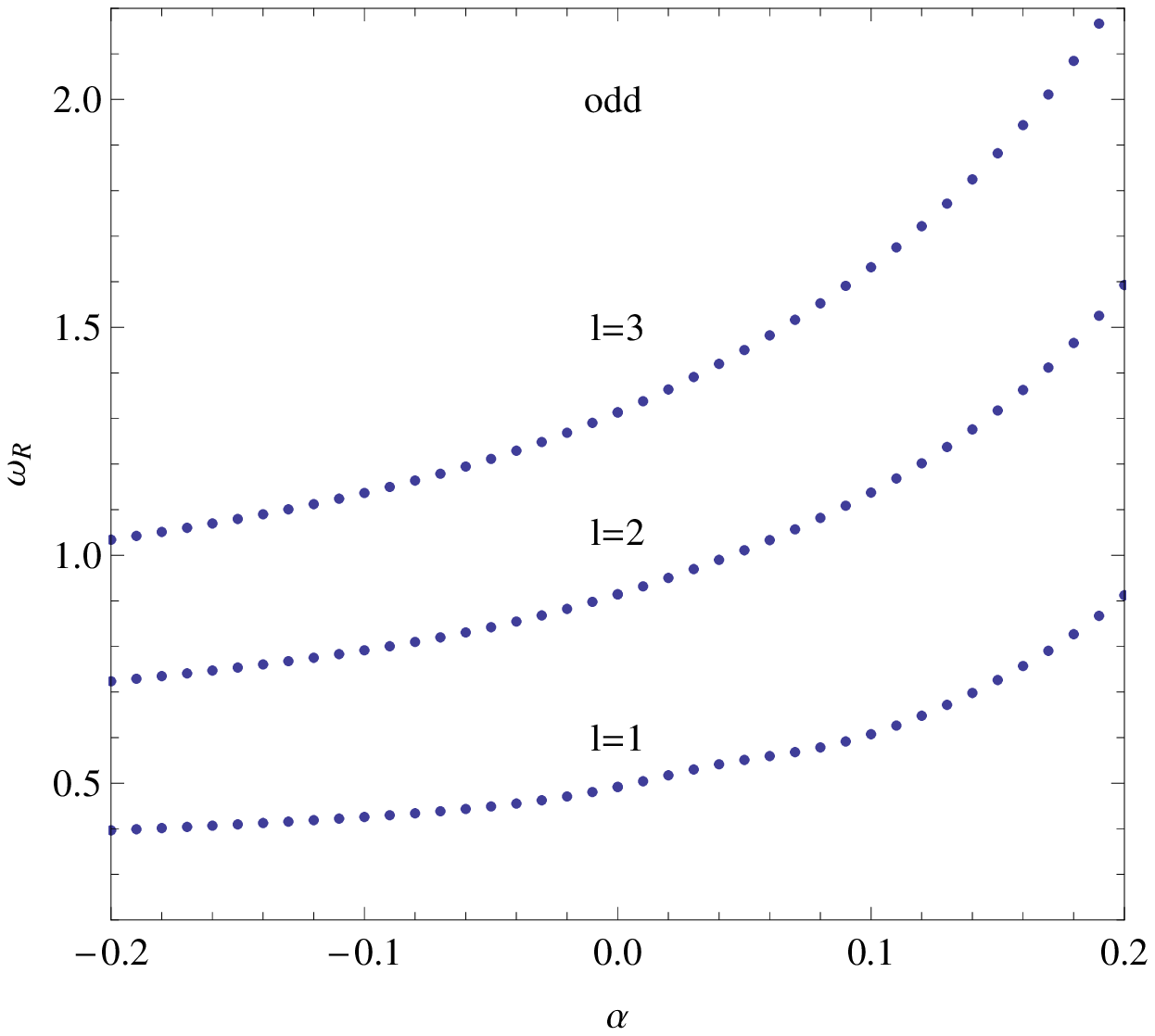}\includegraphics[width=5.6cm]{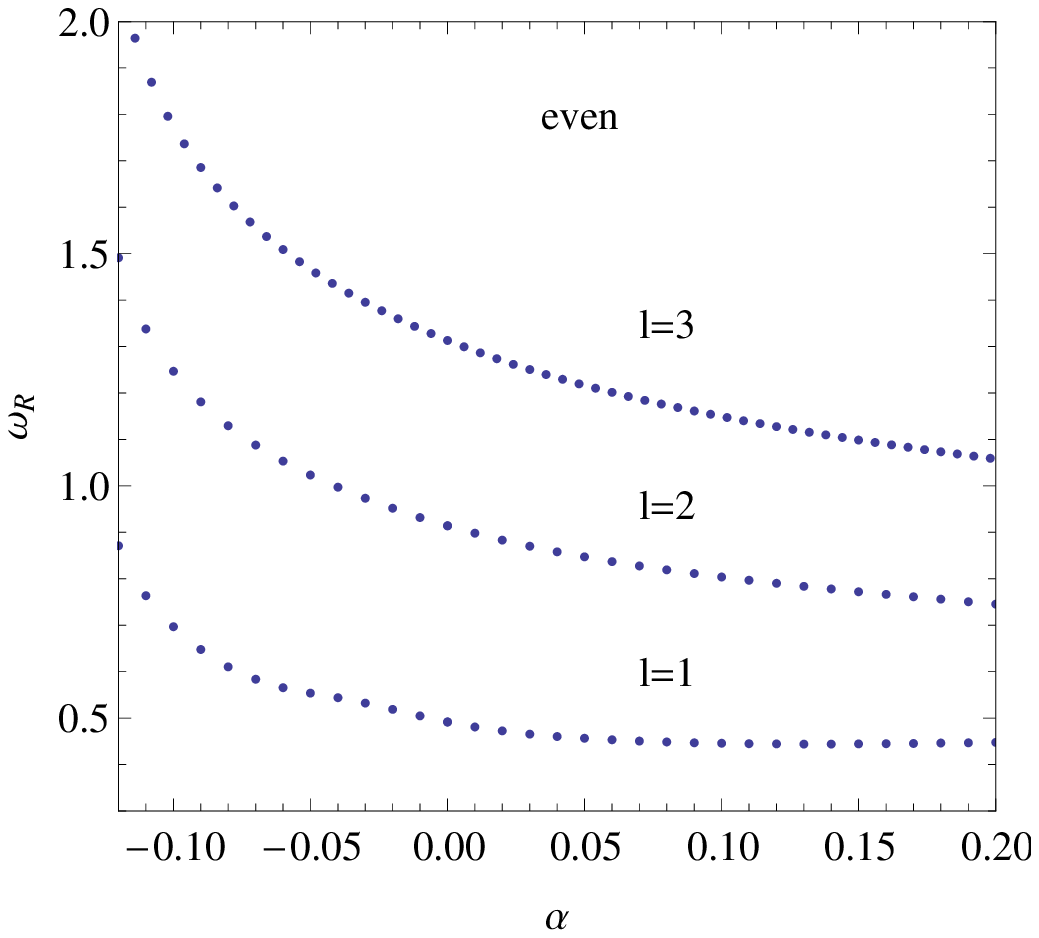}
\caption{Effects of Weyl corrections on the real parts of the
fundamental quasinormal modes of electromagnetic perturbation with
the odd parity(the left) or the even parity (the right)  in the
Schwarzschild black hole spacetime. We set $2M=1$.}
\end{center}
\end{figure}
\begin{figure}[ht]
\begin{center}
\includegraphics[width=5.5cm]{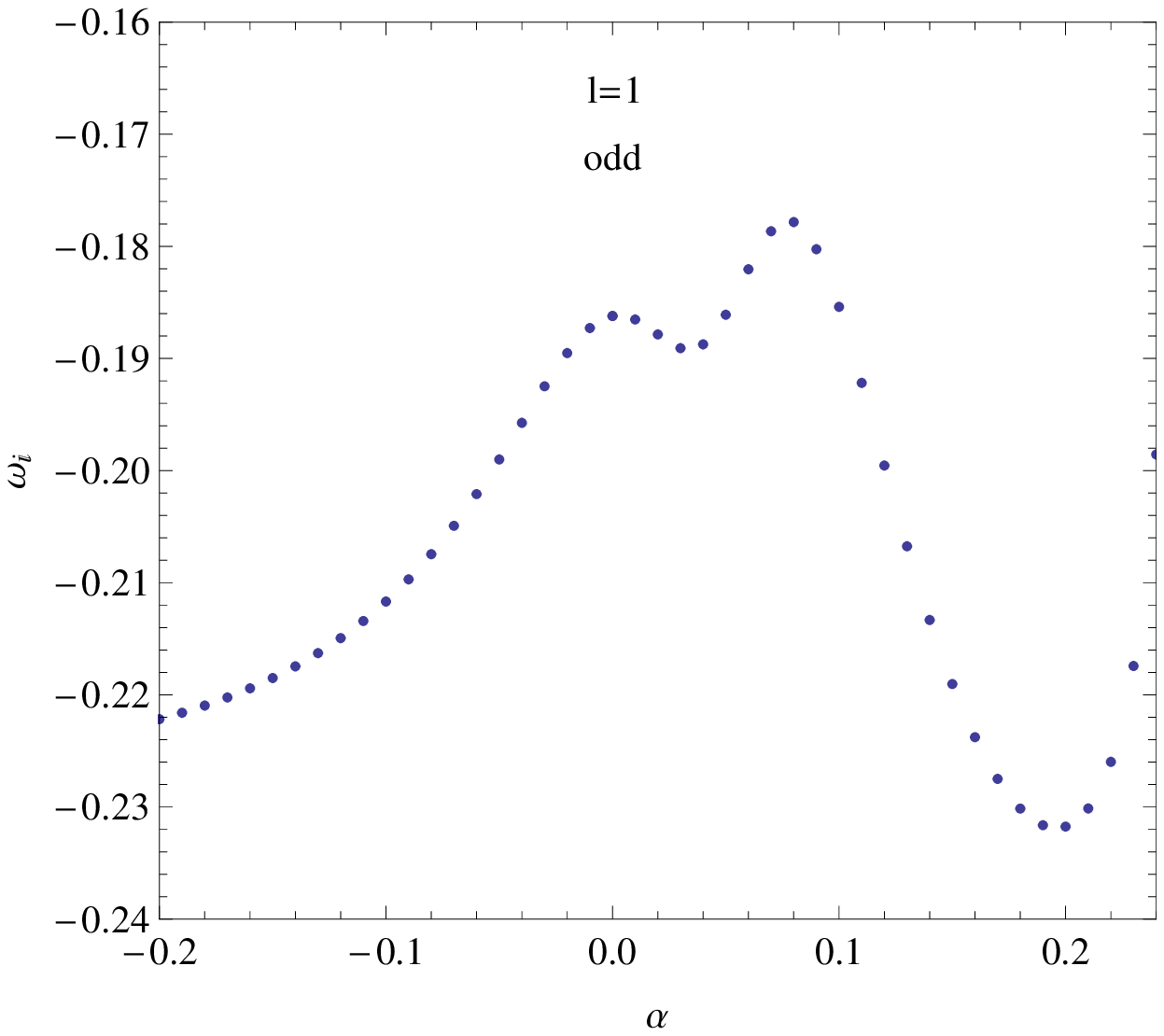}
\includegraphics[width=5.5cm]{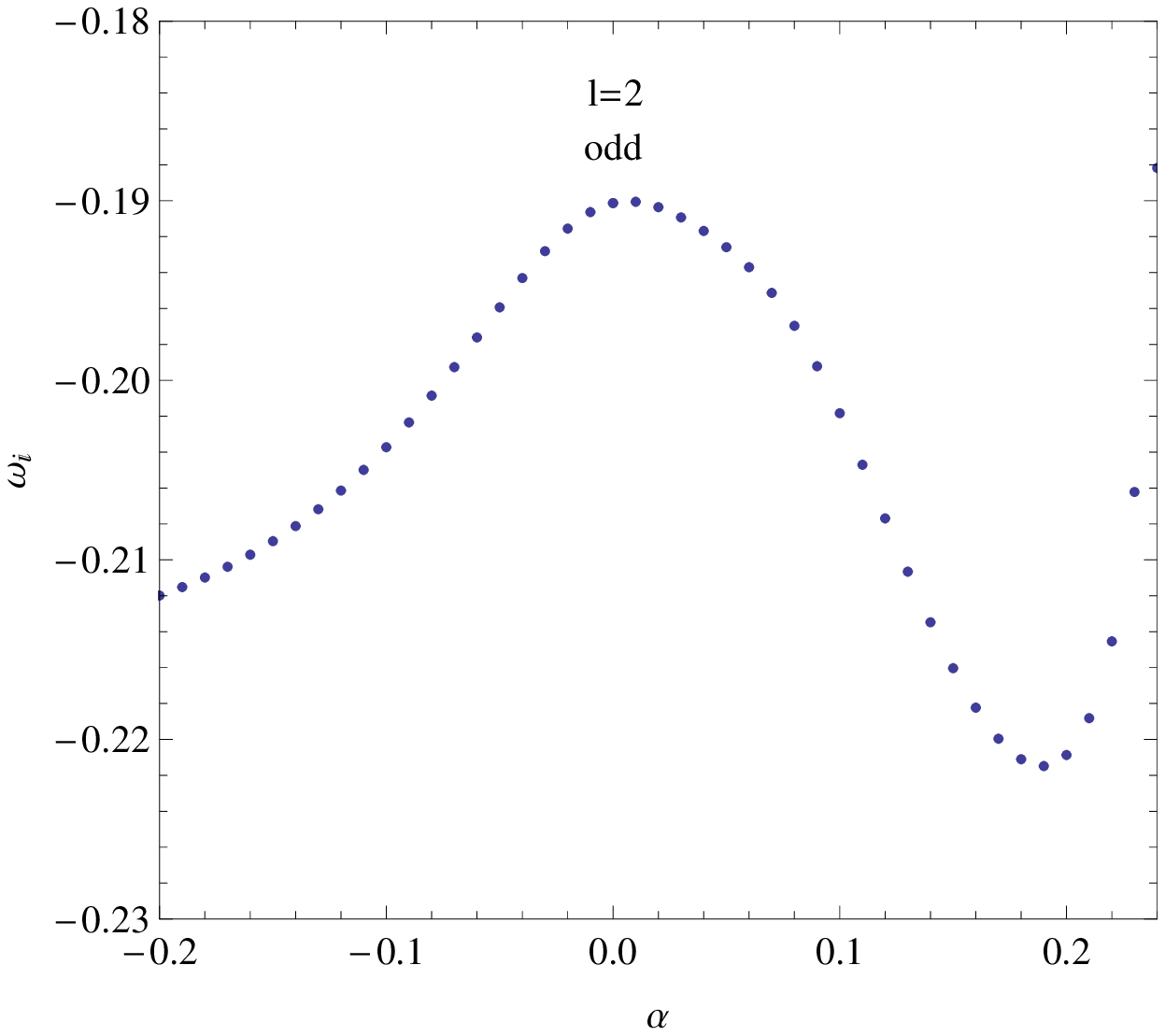}\includegraphics[width=5.5cm]{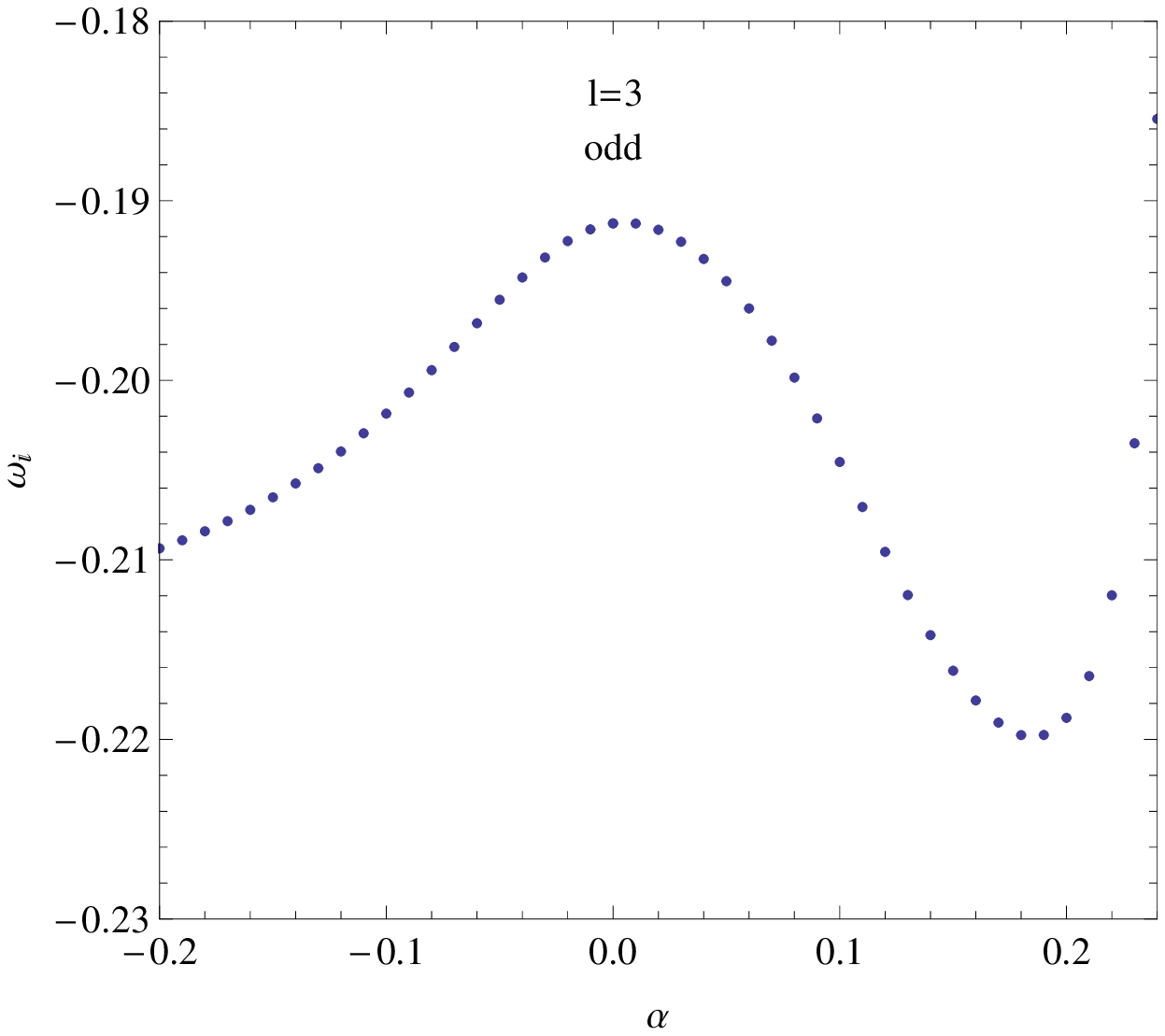}\\
\includegraphics[width=5.5cm]{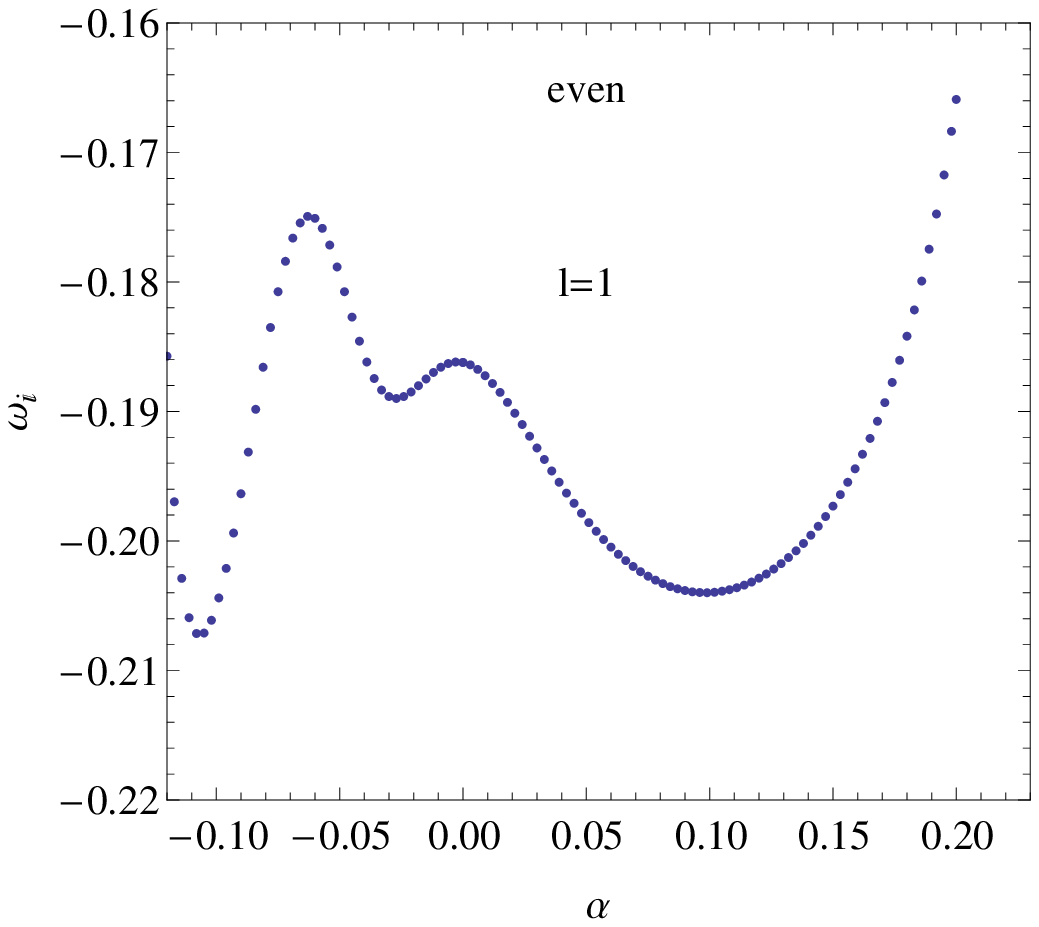}
\includegraphics[width=5.5cm]{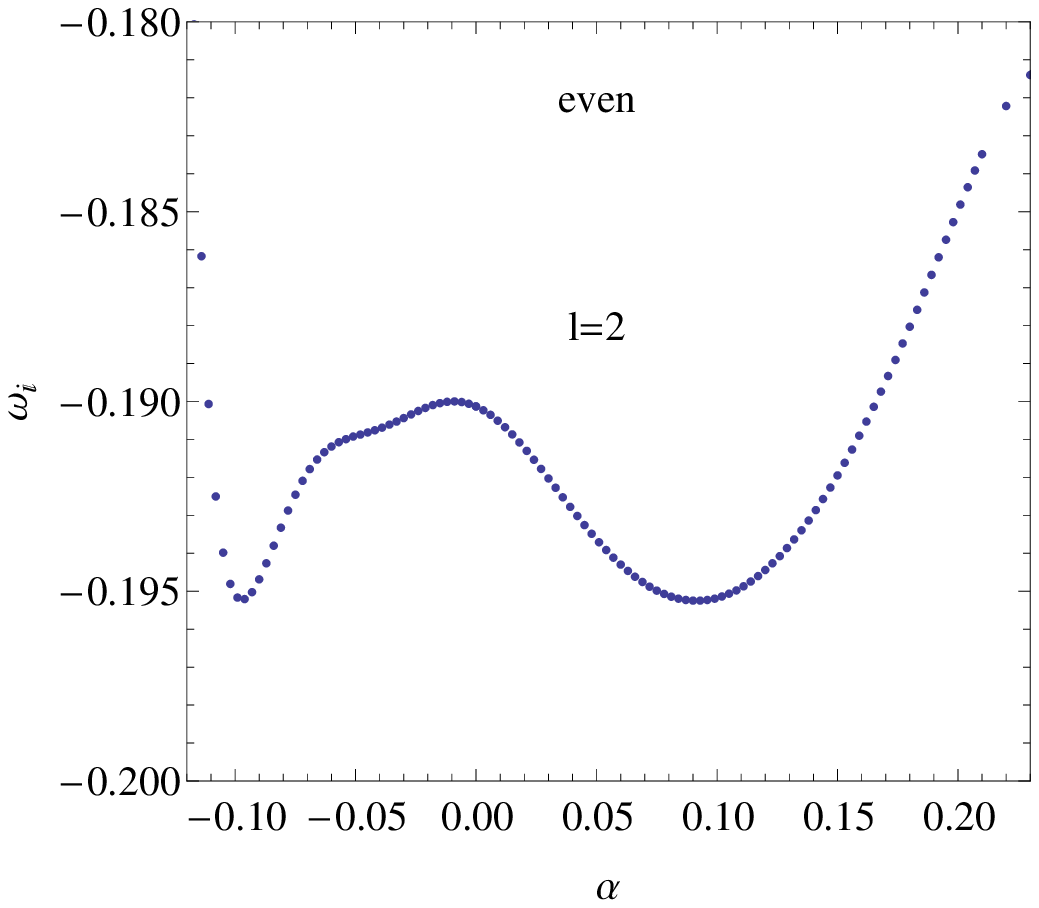}\includegraphics[width=5.5cm]{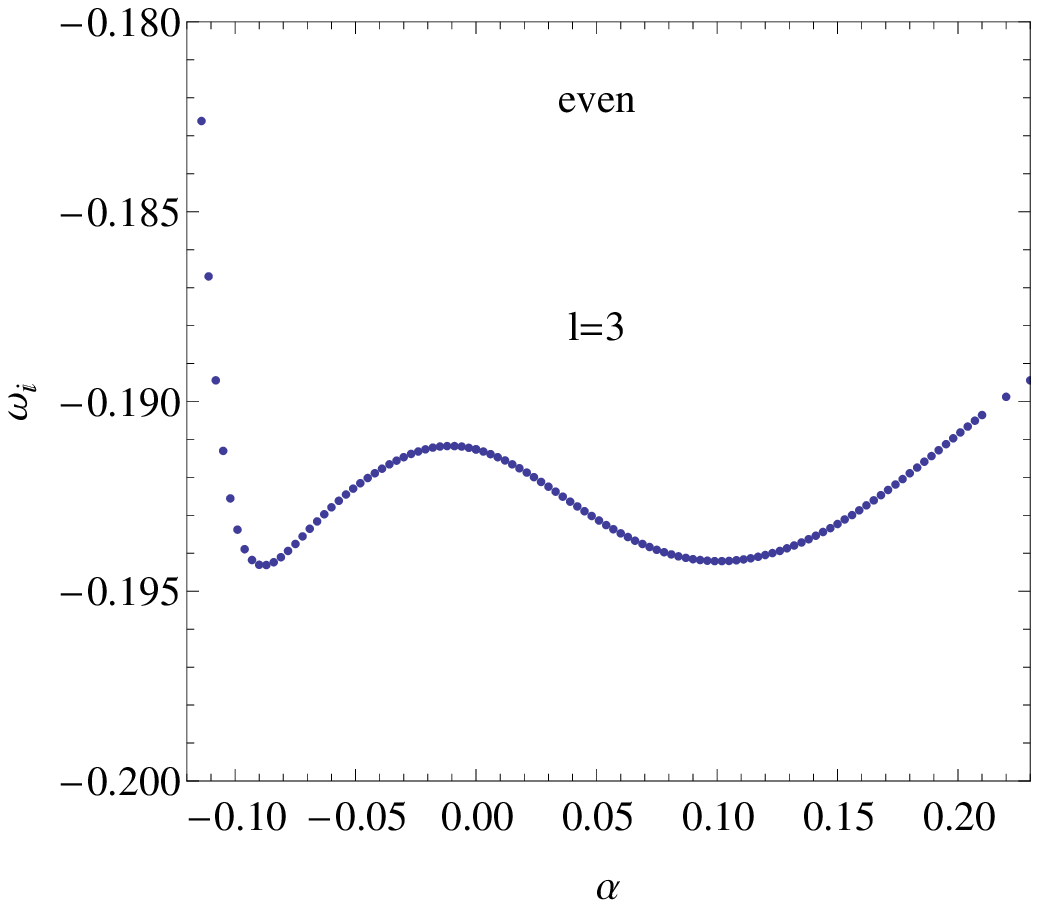}
\caption{Effects of Weyl corrections on the imaginary parts of the
fundamental quasinormal modes of electromagnetic perturbation with
the odd parity(the top row) or the even parity (the bottom row)  in
the Schwarzschild black hole spacetime. We set $2M=1$.}
\end{center}
\end{figure}

Now we are in a position to study the dynamical evolution of the
electromagnetic perturbation with Weyl corrections in time domain \cite{Gundlach:1994}
and examine the stability of the Schwarzschild black hole in this cases. Making use of the light-cone variables $u=t-r_*$ and $v=t+r_*$, one can find
that the wave equation
\begin{eqnarray}
-\frac{\partial^2\psi}{\partial t^2}+\frac{\partial^2\psi}{\partial
r_*^2}=V(r)\psi,
\end{eqnarray}
can be rewritten as
 \begin{eqnarray}
4\frac{\partial^2\psi}{\partial u\partial
v}+V(r)\psi=0.\label{wbes1}
\end{eqnarray}
It is well known that the two-dimensional wave equation (\ref{wbes1}) can be integrated numerically by using the finite difference method suggested in
\cite{Gundlach:1994}. From Taylor's theorem, we can find that the wave equation (\ref{wbes1}) can be
discretized as
\begin{eqnarray}
\psi_N=\psi_E+\psi_W-\psi_S-\delta u\delta v
V(\frac{v_N+v_W-u_N-u_E}{4})\frac{\psi_W+\psi_E}{8}+O(\epsilon^4)=0.\label{wbes2}
\end{eqnarray}
Here we have used the following definitions for the points: $N$:
$(u+\delta u, v+\delta v)$, $W$: $(u + \delta u, v)$, $E$: $(u, v +
\delta v)$ and $S$: $(u, v)$. The parameter $\epsilon$ is an overall
grid scalar factor, so that $\delta u\sim\delta v\sim\epsilon$.
As in \cite{Gundlach:1994}, we can set $\psi(u, v=v_0)=0$ and use a
Gaussian pulse as an initial perturbation, centered on $v_c$ and
with width $\sigma$ on $u=u_0$ as
\begin{eqnarray}
\psi(u=u_0,v)=e^{-\frac{(v-v_c)^2}{2\sigma^2}}.\label{gauss}
\end{eqnarray}
\begin{figure}[ht]
\begin{center}
\includegraphics[width=5.5cm]{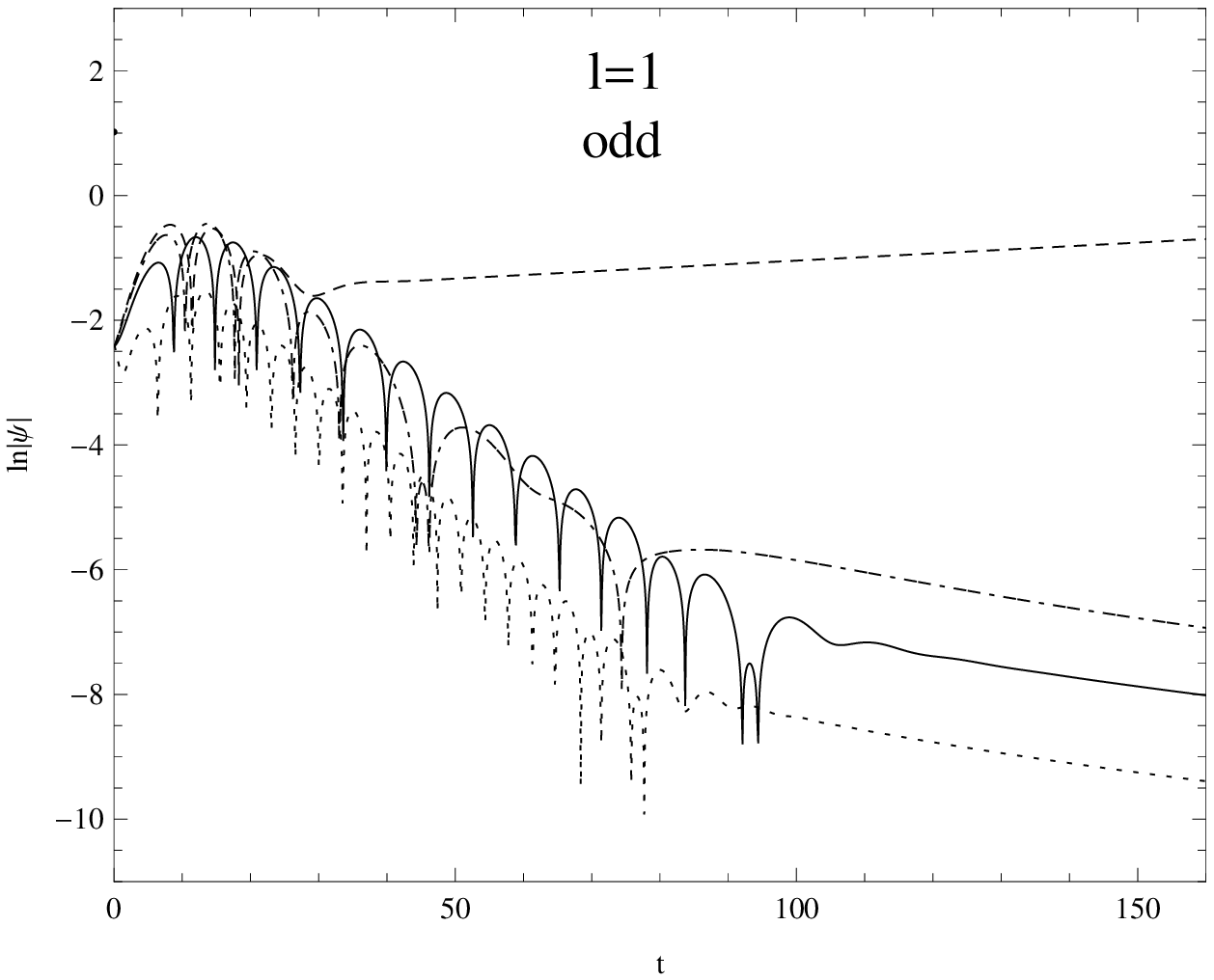}\includegraphics[width=5.5cm]{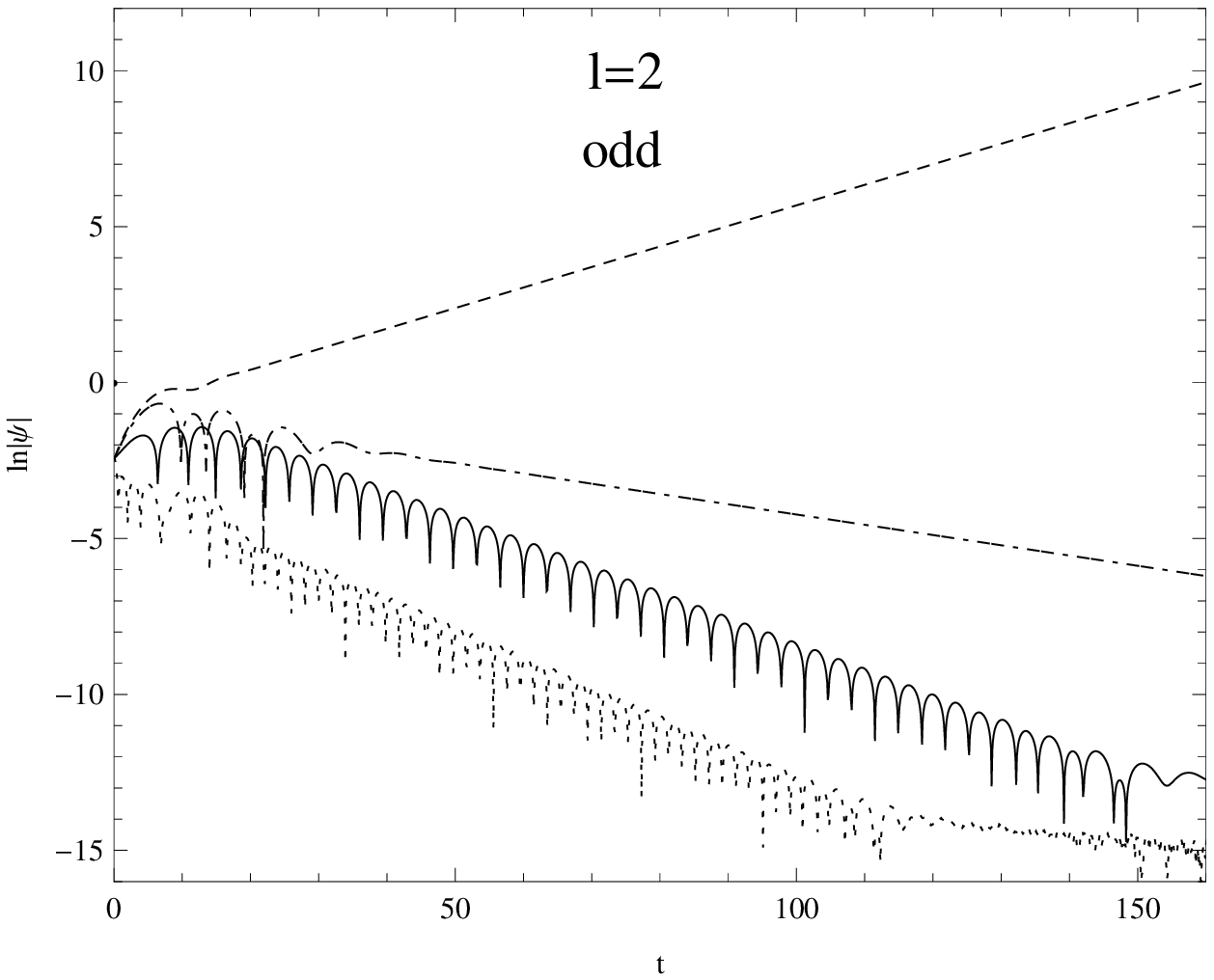}
\includegraphics[width=5.5cm]{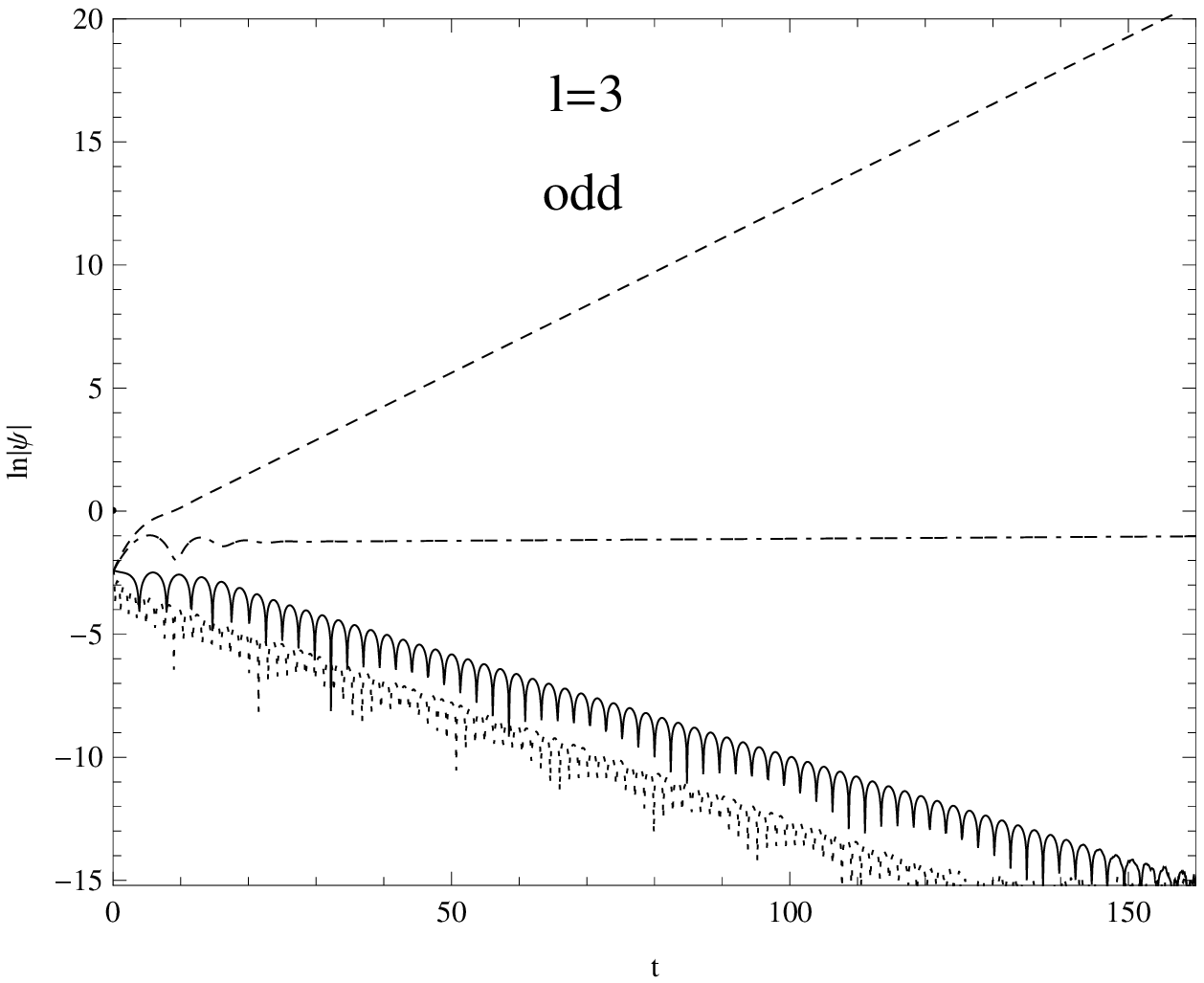}
\caption{The dynamical evolution of an electromagnetic perturbation with odd parity in the background of a Schwarzschild black hole spacetime. The figures from left to right are corresponding to $l=1$, $2$ and $3$. The dotted, solid, dash-dotted and dashed lines are corresponding to the cases with $\alpha=0.2,~0,~-0.2,~-0.3$, respectively. We set $2M=1$.The constants in the Gauss pulse (\ref{gauss}) $v_c=10$ and $\sigma=3$.}
\end{center}
\end{figure}
\begin{figure}[ht]
\begin{center}
\includegraphics[width=5.5cm]{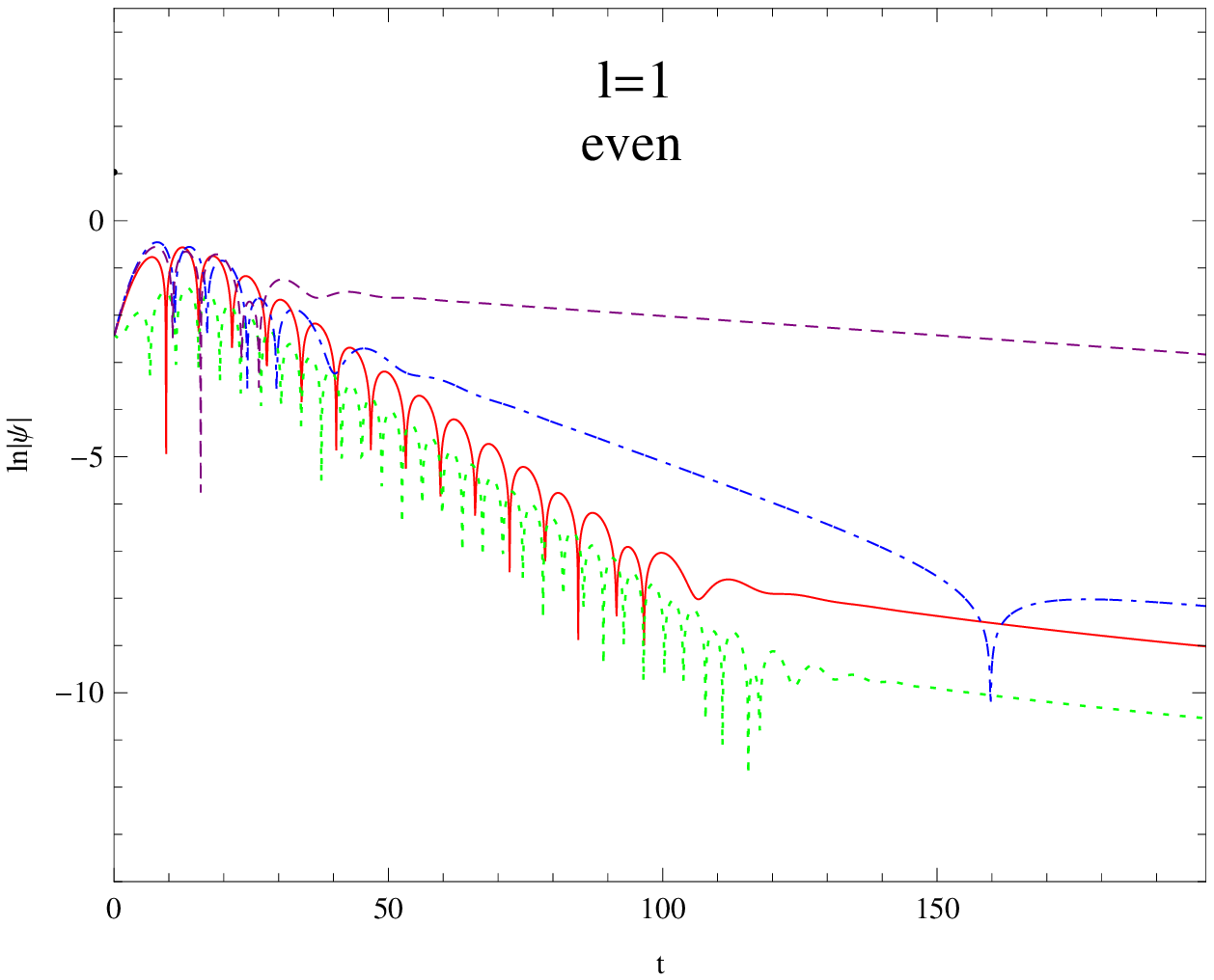}\includegraphics[width=5.5cm]{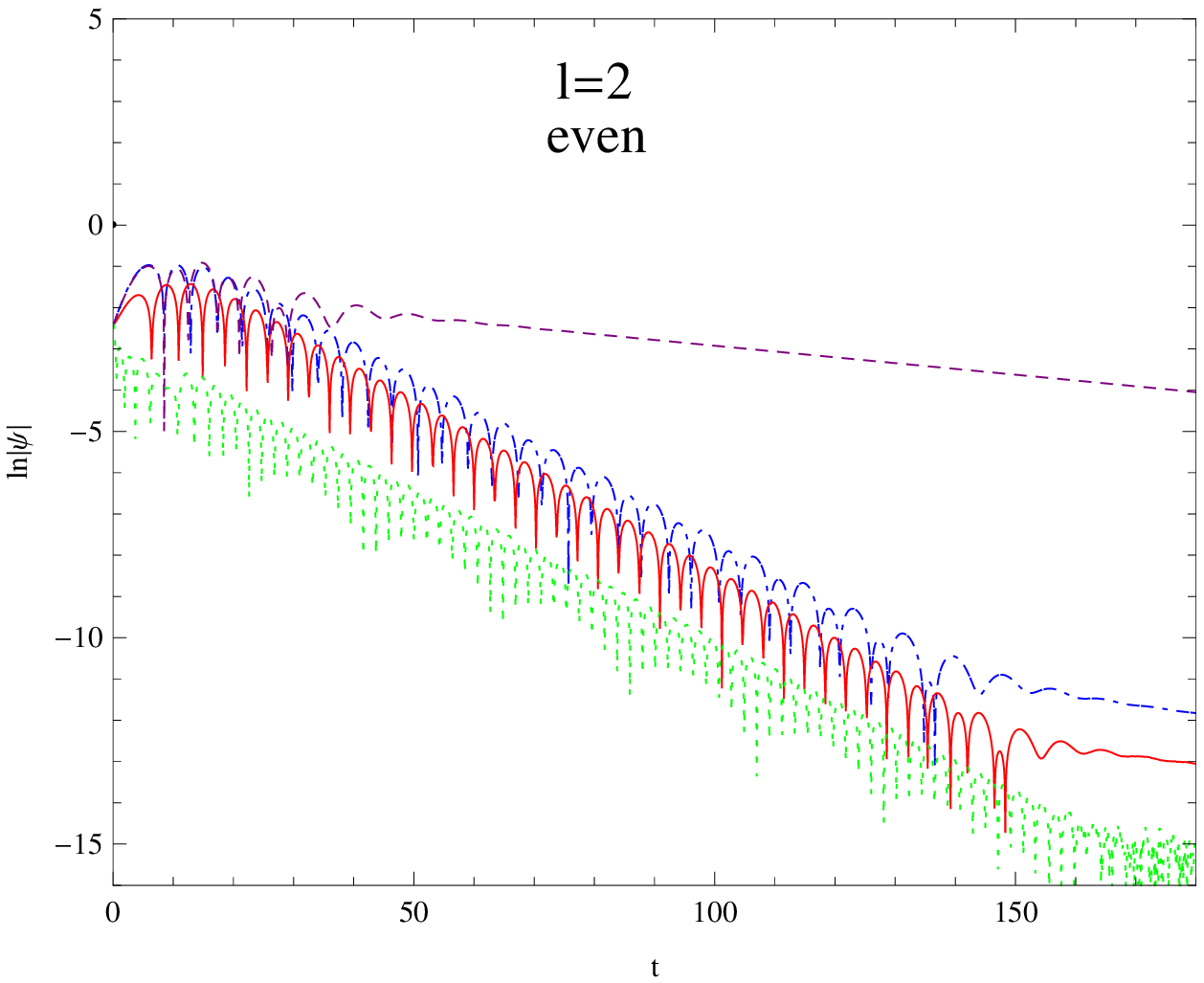}
\includegraphics[width=5.5cm]{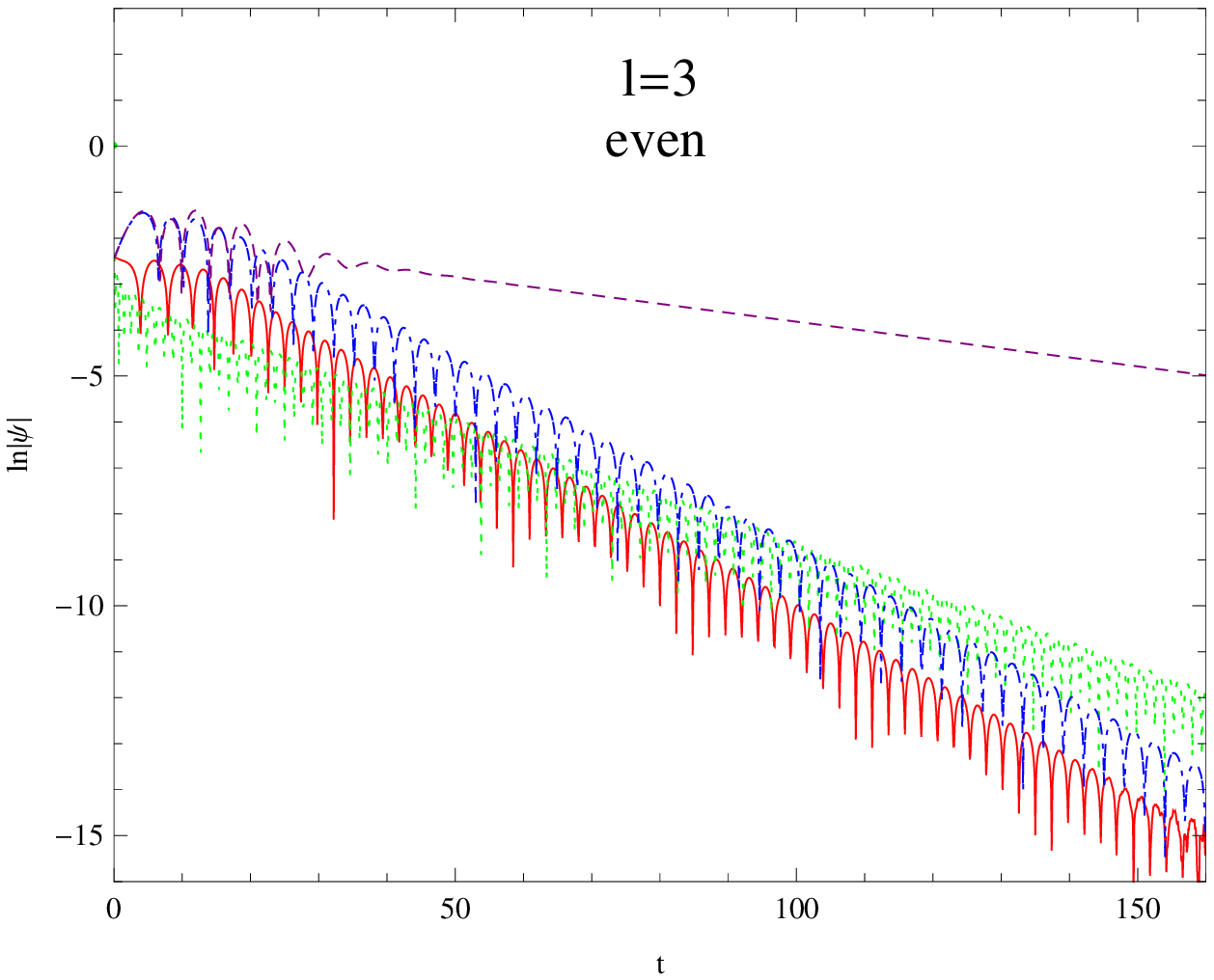}
\caption{The dynamical evolution of an electromagnetic perturbation with even parity in the background of a Schwarzschild black hole spacetime. The figures from left to right are corresponding to $l=1$, $2$ and $3$. The dotted, solid, dash-dotted and dashed lines are corresponding to the cases with $\alpha=-0.12,~0,~0.20,~0.24$, respectively. We set $2M=1$.The constants in the Gauss pulse (\ref{gauss}) $v_c=10$ and $\sigma=3$.}
\end{center}
\end{figure}
In figs. (5) and (6), we present the dynamical evolution of an electromagnetic perturbation with Weyl correction in the background of a
Schwarzschild black hole. Our result show that the effects of Weyl correction on the dynamical evolution of the odd parity electromagnetic perturbation are different from those of the even parity electromagnetic one.
For the electromagnetic perturbation with odd parity, one can that as $\alpha>0$ the decay of the electromagnetic perturbation with the Weyl correction is similar to that of the electromagnetic one without Weyl correction, which
indicates that the black hole is stable. It is understandable by a fact that the effective potential $V(r)$ is positive
definite, which is shown in fig.(1). When $\alpha<0$, we find that
the electromagnetic field grows with exponential rate if the coupling constant
$\alpha$ is smaller than the critical value $\alpha_c$. It means that
the instability occurs in this case. The main reason is that for
the odd parity electromagnetic perturbation with negative coupling constant $\alpha$ the large absolute value of $\alpha$ drops down the peak of the potential barrier and increases the negative gap near the black hole horizon
so that the potential could be non-positive definite. In the
instability region, we can find that for the larger $|\alpha|$ the instability growth appears more early and the growth rate becomes stronger.
For the electromagnetic perturbation with even parity, we find that the electromagnetic field always decays in the allowed range of $\alpha$. Although in the effective potential $V(r)_{even}$ the negative gap appears near the black hole horizon as $\alpha>0$ and increases with $\alpha$, the negative gap is very small in this case, which is not enough to yield the instability to be triggered.

In fig.(7), we plotted the change of the threshold value $\alpha_c$
with $l$ for the odd parity electromagnetic perturbation with Weyl corrections, and found that the threshold value can be fitted
best by the function
\begin{eqnarray}
\alpha_c\simeq a\sqrt{\frac{l}{l+1}}+b,\label{ncs}
\end{eqnarray}
where the coefficients $a$ and $b$ are numerical constants with dimensions of length-squared and their values are $a=0.5534$ and $b=-0.6785$.
\begin{figure}[ht]
\begin{center}
\includegraphics[width=7cm]{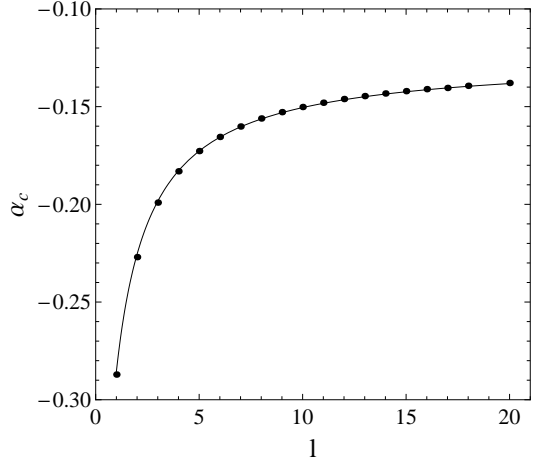}
\caption{The change of the threshold value $\alpha_c$ with the multipole number $l$. The points $l=1\sim 20$ are fitted by the function $\alpha_c=a~\sqrt{\frac{l}{l+1}}+b$, where the coefficients $a$ and $b$ are numerical constants with dimensions of length-squared and their values are $a=0.5534$ and $b=-0.6785$.}
\end{center}
\end{figure}
It is easy to obtain that the threshold value $\alpha_c$ is negative and increases with the multipole number $l$. This means that for the
higher $l$ we need the smaller Weyl corrections at which instability happens.
From Eq. (\ref{ncs}), we can obtain that in the limit $l\rightarrow\infty$
the threshold value $\alpha_c\rightarrow a+b=-0.1251$. It could be explained by a fact that in this limit the effective
potential (\ref{evodsch}) has the form
\begin{eqnarray}
V(r)|_{l\rightarrow\infty}=f\frac{l(l+1)}{r^2}\bigg(\frac{r^3+16\alpha
M}{r^3-8\alpha M}\bigg),\label{evl}
\end{eqnarray}
which leads to the integration \cite{Gleiser:2005},
\begin{eqnarray}
\int^{\infty}_{r_+}\frac{V(r)|_{l\rightarrow\infty}}{f}dr=\int^{\infty}_{r_+}\bigg(\frac{r^3+16\alpha
M}{r^3-8\alpha M}\bigg)\frac{l(l+1)}{r^2}dr,
\end{eqnarray}
is positive definite as $\alpha<-r^3_+/(16M)$. It implies that the
threshold value has a form $\alpha_c=-M^2/2$ as
$l\rightarrow\infty$, which is consistent with the form of the
numerical constant $a+b$ obtained in the previous numerical
calculation.

\section{summary}
In this paper, we present firstly the master equation of an electromagnetic perturbation with Weyl correction in the four-dimensional black hole spacetime and find that the presence of Weyl corrections makes that the master equation of the odd parity electromagnetic perturbation is different from that of the even parity one, which is quite different from that of the usual electromagnetic perturbation without Weyl correction in the four-dimensional spacetime where the master equation is independent of the parity of the electromagnetic field. And then we have investigated numerically the dynamical evolution of an electromagnetic perturbation with Weyl correction in the background of
a four-dimensional Schwarzschild black hole spacetime. Our results show that the Weyl correction modifies the standard results of the wave dynamics for the electromagnetic perturbation. Due to the presence of Weyl corrections, the dynamical properties of the electromagnetic perturbation depend not only on the Weyl correction parameter $\alpha$, but also on the parity of the electromagnetic field. With the increase of $\alpha$, the real part of the fundamental quasinormal frequencies for fixed $l$ increases for the odd parity electromagnetic perturbation and decreases for the event parity electromagnetic one. The changes of the imaginary parts with
$\alpha$ are more complicated. Moreover, we find that
the odd parity electromagnetic perturbation grows with exponential rate if the coupling constant
$\alpha$ is smaller than the negative critical value $\alpha_c$.  This means that the instability occurs in this case. In the
instability region, we can find that for the smaller $\alpha$ the instability growth appears more early and the growth rate becomes stronger.
For the electromagnetic perturbation with even parity, we find that the electromagnetic field always decays in the allowed range of $\alpha$. Although in the effective potential $V(r)_{even}$ the negative gap appears near the black hole horizon as $\alpha>0$, the negative gap is very small and is not enough to yield the occurrence of instability in this case.

Furthermore, we find that the threshold value can
be fitted best by the function $\alpha_c\simeq a\sqrt{\frac{l}{l+1}}+b$ with two numerical constants (with dimensions of length-squared) $a, b$. These rich
dynamical properties of the electromagnetic perturbation with Weyl correction, at least in principle, may provide a possibility  to detect whether there exists Weyl correction to electromagnetic field or not in the future astronomical observations, such as the future
space-based detector LISA etc. It
would be of interest to generalize our study to other black hole
spacetimes, such as rotating black holes etc. Work in this direction
will be reported in the future.

\section{\bf Acknowledgments}

This work was  partially supported by the National Natural Science
Foundation of China under Grant No.11275065, the NCET under Grant
No.10-0165, the PCSIRT under Grant No. IRT0964,  the Hunan
Provincial Natural Science Foundation of China (11JJ7001) and the
construct program of key disciplines in Hunan Province. J. Jing's
work was partially supported by the National Natural Science
Foundation of China under Grant Nos. 11175065, 10935013; 973 Program
Grant No. 2010CB833004.

\end{document}